\documentclass[useAMS,usenatbib]{mn2e}
\usepackage{graphicx,psfig,epsfig}

\def\lsim{\,\lower2truept\hbox{${<\atop\hbox{\raise4truept\hbox{$\sim$}}}$}\,}
\def\gsim{\,\lower2truept\hbox{${> \atop\hbox{\raise4truept\hbox{$\sim$}}}$}\,}

\title[Forecast B-modes detection at large scales]{Forecast B-modes detection at large scales in presence of noise and foregrounds}
\author[Bonaldi A. and Ricciardi S.]{Bonaldi A. $^{1}$ and Ricciardi S. $^{2}$\\
$^{1}$INAF, Osservatorio Astronomico di Padova, Vicolo dellOsservatorio 5, I-35122 padova, Italy \\
$^{2}$INAF/IASF, Sezione di Bologna, Via Gobetti 101, I-40129 Bologna, Italy}
\begin{document}

\maketitle

\begin{abstract}
We investigate the detectability of the primordial CMB polarization B-mode power spectrum on large scales in the presence of instrumental noise and realistic foreground contamination. We have worked out a method to estimate the errors on component separation and to propagate them up to the power spectrum estimation. The performances of our method are illustrated by applying it to the instrumental specifications of the {\sc Planck} satellite and to the proposed configuration for the next generation CMB polarization  experiment COrE. We demonstrate that a proper component separation step is required in order achieve the detection of B-modes on large scales and that the final sensitivity  to B-modes of a given experiment is determined by a delicate balance between noise level and residual foregrounds, which depend on the set of frequencies exploited in the CMB reconstruction, on the signal-to-noise of each frequency map, and on our ability to correctly model the spectral behavior of the foreground components. We have produced a flexible software tool that allows the comparison of performances on B-mode detection of different instrumental specifications (choice of frequencies, noise level at each frequency, etc.) as well as of different proposed approaches to component separation.
\end{abstract}

\section{Introduction}
Primordial B-mode (curl component) polarization of the CMB provides a unique opportunity to detect the imprint of the primordial gravitational waves predicted by the inflationary paradigm. Measuring the amplitude of these tensor perturbations would fix the energy scale of inflation and its potential and would provide a powerful constraint on a broad class of inflationary models. Moreover, the confirmation of inflation and the determination of the inflationary potential would have profound
implications, by providing a direct observational link with the physics of the early universe.

The bulk of the statistical information on inflationary B-modes is concentrated in two features of the power spectrum: the reionization bump at multipoles $\ell \sim 2-10$  and the main bump at multipoles $\ell \sim 20-100$. Given that most of the signal lies on large angular scales (larger than $\sim 1^\circ$), full-sky surveys, and thus satellite experiments, are required to limit cosmic variance.

When designing a satellite experiment targeted to B-mode detection on large scales it is mandatory to consider, together with the signal-to-noise ratio, also the issues related to foreground contamination and component separation.
In fact, as demonstrated by the WMAP data (Gold et al. 2009), the cleaning of CMB E-mode from polarized foregrounds is still a manageable problem. But polarized foregrounds are expected to dominate by a wide margin over the CMB B-mode because foreground E- and B-modes have similar amplitudes while the CMB B-mode is far weaker than the E-mode.

The constraints in detecting primordial B-modes due to foreground contamination and residuals from foreground subtraction have been investigated by several authors. Many papers include a modeling of the residuals in the Fisher matrix (see. e.g., Tucci et al. 2005, Amarie et al. 2005, Verde et al. 2006) but generally bypass any realistic component separation approach and thus cannot properly deal with component separation errors. Other analyses focussed on a certain component separation method and evaluated its performances through simulations tailored for a specific experiment.  For example, Betoule et al. (2009) and  Efstathiou et al. (2009) propose different strategies (a spectral matching component separation method and an internal template fitting method, respectively) targeted to B-mode detection for the {\sc Planck} experiment. More recently, Stivoli et al. (2010) assessed the capabilities of a parametric component separation method to detect B-modes for suborbital experiments.

The purpose of our work is different: while understanding the need of performing detailed simulations once the experimental configuration is established, we want to provide a fast, agile and flexible tool \footnote{the software could be available upon request} to forecast the detectability of B-modes  allowing us to easily change both the instrumental specifications and the CMB reconstruction strategy. Such a tool can be usually exploited to help optimizing the instrumental design and the data reduction pipeline for future CMB polarization experiments.

%agile veloce
%With this work we want to provide a general and flexible tool able to forecast the detectability of B-modes given certain instrumental specifications in presence of component separation. We target to large scales and thus mainly to full-sky experiments.
The different components are reconstructed as a suitable linear mixture of the data. Different choices for the linear mixture operator are available, and we carry out a comparison of their performances %The possible exploitations of our tool are many, from the instrumental design to the definition of the data reduction pipeline for a given experiment.
The forecast, targeted to large scales, on the B-mode detectability is performed at the power spectrum level and includes the computation of the final power spectra of noise and foreground residuals on the CMB map. We explicitly account for errors in the frequency scalings assumed for the foreground components. On the other hand, we do not account for errors due to actual power spectrum estimation (leakage effects) nor for instrumental systematics. The former should be subdominant for the almost full-sky coverage considered here, especially when exploiting a power spectrum estimation method optimized for low multipoles (e.g. Gruppuso et al. 2009). Instrumental systematics  obviously depend on the particular instrument and thus cannot be included in our general treatment.

The outline of our paper is the following: in \S\,\ref{sec:due} we describe our models for the data and the component separation process.  In \S\,\ref{sec:errbar} we derive analytical expressions for the noise and residual foreground contributions to the errors in the $BB$  power spectrum. In \S\,\ref{sec:quattro} we apply our method to the specifications of the {\sc Planck} satellite and of a proposed CMB polarization experiment (COrE white paper in preparation). Our results are presented in \S\,\ref{sec:results} and our main conclusions are summarized in \S\,\ref{sec:conclusions}.

\section{Statement of the problem}
\label{sec:due}
\subsection{Data model}
The microwave sky contains, besides the CMB, several foreground components, both diffuse and compact. For our analysis, which is focused on large and intermediate scales, we will consider only diffuse foregrounds. The main diffuse polarized foregrounds are Galactic synchrotron and thermal dust (the free-free emission is unpolarized and the anomalous dust emission is also expected to be essentially unpolarized).

The synchrotron component dominates at the lower frequencies. Its spectral behavior in antenna temperature can be modeled as a power law:
\begin{equation}\label{scaling_synchro}
T_{\rm A,synch}(\nu)\propto \nu^{-\beta_s} \label{syn}\ ,
\end{equation}
where the synchrotron spectral index $\beta_s$ can vary in the sky in the range $2.5<\beta_s<3.5$ (Gold et al. 2009). The spectral behaviour of thermal dust emission, that takes over at high frequencies, follows approximately a grey-body law:
\begin{equation}\label{scaling_dust}
T_{\rm A,dust} (\nu)\propto \frac{\nu^{\beta_d+1}}{\exp
(h\nu/kT_{\rm dust})-1}. \label{dust}
\end{equation}
Both $\beta_d$ and $T_d$ are spatially varying around $\beta_d \sim 1.7$ and $T_d \sim 18\,$K. The polarized CMB signal has a blackbody spectrum:
\begin{equation}\label{scaling_CMB}
T_{\rm A,CMB}(\nu)\propto\frac{(h\nu/kT_{\rm CMB})^2\exp (h\nu/kT_{\rm CMB})}
{(\exp (h\nu/kT_{\rm CMB})-1)^2} \
\end{equation}
with $T_{\rm CMB}=2.73$.

The sky radiation, $\tilde{x}$, from the direction $\hat{r}$ at the frequency $\nu$ is  the superposition of signals coming from $N_c$ different physical processes $\tilde{s}_j$:
\begin{equation}
\tilde{x}(\hat{r},\nu)=\sum_{j=1}^{N_c}\tilde{s}_j(\hat{r},\nu).
\end{equation}
The instrument integrates the signal over frequency in each of its $N_d$ channels, convolves it with a certain kernel, representing the spatial response function, and adds noise. The measurement yielded by the generic channel centered at the frequency $\nu$, with beam axis in the direction $\hat{r}$ is:
\begin{equation}
x_\nu(\hat{r})=\!\!\int\!\! B(\hat{r}-\hat{r'},\nu')\sum_{j=1}^{N_c}t_\nu(\nu')
\tilde{s}_j(\hat{r',\nu'}) dr' d\nu'+n_\nu(\hat{r})\label{m0},
\end{equation}
where $ B(\hat{r}-\hat{r'},\nu')$ is the (azimuthally-symmetric) beam, $t_\nu(\nu')$ is the frequency response of the channel and $n_\nu(\hat{r})$ is the noise map.  The data model in eq.~(\ref{m0}) can be simplified under the following
assumptions:
\begin{enumerate}
\item Each source signal, $\tilde{s}_j(\hat{r},\nu)$, is a separable function of direction and frequency at least within limited sky patches;
%\begin{equation}
%\tilde{s}_j(\hat{r},\nu)=\bar{s}_j(\hat{r})f_j(\nu)\label{3}
%\end{equation}
\item $B(\hat{r},\nu)$ is independent of frequency within the passband of each channel;
\item $B(\hat{r},\nu)$ is the same for all the channels.
%\item the beam patterns of the telescope are the same for
%all the measurement channels
\end{enumerate}
The first two are reasonable approximations. To meet the third one we need to smooth the data to the lowest resolution we are dealing with. This is not a problem in the present context, since we are interested in large angular scales.

Under these assumptions, the \emph{linear mixture} data model applies. In vector form, for each direction of the sky (each pixel) we may write:
\begin{equation}\label{datamodel}
\bmath{x}=\bmath{\sf{H}}\bmath{s}+\bmath{n}.
\end{equation}
where $\bmath{x}$ and $\bmath{n}$ are $N_d$-vectors containing data and instrumental noise respectively,  $\bmath{s}$ is the $N_c$-vector containing sources, and $\bmath{\sf{H}}$ is the $N_d \times N_c$ mixing matrix, containing the frequency scaling of the components. The spatial variability of the synchrotron and dust spectral indices implies that the mixing matrix $\bmath{\sf{H}}$ is in general different for different pixels.

\subsection{CMB recovery}

If the {linear mixture} data model holds, the components which are mixed in the channel maps $\bmath{x}$ [eq.~(\ref{datamodel})] can be reconstructed as:
%get an estimate  $\bmath{\hat s}$ of all the true components $\bmath{s}$ through:
%
\begin{equation}\label{recon}
\bmath{\hat s}=\bmath{\sf{W}}\bmath{x}.
\end{equation}
where  $\bmath{\hat s}$ is an estimate of the components $\bmath{s}$ and $\bmath{\sf{W}}$ is a $N_c \times N_d$ matrix called \emph{reconstruction matrix}.
We choose to rely on a linear estimator basically because it will allow us to easily deal with the component separation process in the forecast of B-mode detection, as we will see in the next section. In addition this approach is handy for  Monte Carlo simulations needed to accurately control error sources.

We adopt  the so-called Generalized Least Square solution (GLS):
\begin{equation}
\label{gls}
\bmath{\sf{W}}=[\bmath{\sf{\hat H}}^T \bmath{\sf{N}}^{-1}\bmath{\sf{\hat H}}]^{-1}\bmath{\sf{\hat H}}^T\bmath{\sf{N}}^{-1}.
\end{equation}
This requires the noise covariance $\bmath{\sf{N}}$ of the channel maps and an estimate $\bmath{\sf{\hat H}}$ of the mixing matrix $\bmath{\sf{H}}$ that can be obtained exploiting any of the successful component separation methods discussed in the literature (see, e.g., Bonaldi et al. 2006, Eriksen et al. 2006, Stompor et al. 2009).
We stress that this choice is  not completely general, as other reconstruction matrices could be used. The adaptation of our approach to other choices for $\bmath{\sf{W}}$, still in the framework of the linear mixture data model, is however straightforward.

The fact that our reconstruction matrix explicitly contains the estimated mixing matrix will allow us to easily introduce in our forecast also the effect of errors in the mixing matrix estimation (see \S\,\ref{sec:errbar}).

The noise covariance term, $\bmath{\sf{N}}$, in eq.~(\ref{gls}) allow us to perform the  component separation keeping the noise level under control. Nevertheless, every component separation leads to a noise amplification by an amount ultimately depending on the conditioning of the mixing matrix $\bmath{\sf{\hat H}}$, as we will show in \S\,\ref{sec:quattro}. 
Despite this drawback, in the low multipole regime
considered  we can not dispense from the component
separation step (e.g. by resorting to an aggressive
masking of foreground contaminated regions) because the
foreground contamination is more troublesome than the noise amplification
and large areas are essential to restrain cosmic variance.

On the contrary, the use of a suitable mask could be enough to keep foreground contamination under control on smaller scales, where the power spectra of diffuse foregrounds decrease more rapidly than those of CMB and noise. In this case we could completely change strategy and simply recover the CMB as a weighted mean of the channels with weights given by the inverse noise variance. This can be viewed again as a linear combination as in eq.~(\ref{recon}), where the reconstruction matrix $\bmath{\sf{W}}$ has null elements for the foregrounds and inverse noise variance weights for the CMB.  This approach  is complementary to the one previously described: it allows noise reduction but does not separate the components; only one component is reconstructed and considered as CMB.

In this work we will consider both approaches, which will be labeled as ``component separation'' (CS) and ``minimum variance'' (MV). The comparison of the results in \S\,\ref{sec:results} will give us an idea of the multipole ranges where each approach works better.

\begin{table*}
\centering
\begin{minipage}{140mm}
\caption{Instrumental characteristics considered in the present study for the {\sc Planck} and COrE experiments. The rms per pixel is referred to nside 1024 (size $\sim 3.5\,$arcmin)} \label{tab:rms}
\begin{tabular}{llllllllllllllllll}
\hline
{\sc Planck} specifications\footnote{LFI specifications as in Mandolesi et al. (2010), HFI specifications as in Lamarre et al. (2008)}\\
\hline
$\nu\,$(GHz)&30&44&70&100&143&&&217&&&&353&\\
FWHM (arcmin)&33&24&14&9.5&7.1&&&5.0&&&&5.0&\\
%NE$\Delta T$ ($\mu$K RJ$\sqrt(s)$) &116 & 113 & 105&27& 18 &&16& 12&\\
29 m RMS $\Delta T$($\mu$K RJ)&66 & 64 &60 & 20 &10 &&&7.6 &&&& 7.5&\\
\hline
COrE specifications \footnote{ESA M3 call (Dec 2010) available at \emph{http://www.core-mission.org}} \\
\hline
$\nu\,$(GHz)&&45&75&105&135&165&195&225&255&285&315&375&435&555&675&795\\
FWHM (arcmin)&&23&14&10&7.8&6.4&5.4&4.7&4.1&3.7&3.3&2.8&2.4&1.9&1.6&1.3\\
%NE$\Delta T$ ($\mu$K RJ$\sqrt(s)$)&&&  & &  &  & &  & \\
24 m RMS $\Delta T$($\mu$K RJ))&&1.4&0.7&0.6&0.5&0.4&0.3&0.2&0.4&0.5&1.1&3.6&0.9&1.2&1.2&1.3\\
\hline
\end{tabular}
\end{minipage}
\end{table*}

\section{Forecast of B-mode detection}\label{sec:errbar}
Our forecast is done at the power spectrum level and accounts for both noise errors and foreground residuals:
\begin{equation}
\label{total}
\Delta \bmath{C}_{\ell}=\Delta \bmath{C}_{\ell,{\rm noise}}+\Delta \bmath{C}_{\ell,{\rm foreg}}.
\end{equation}
In this section we formalize the estimate of the two contributions $\Delta \bmath{C}_{\ell,{\rm noise}}$ and $\Delta \bmath{C}_{\ell,{\rm foreg}}$ to the error on the power spectrum, exploiting the model of the sky [eq.~(\ref{datamodel})] and of the component separation process [eq.~(\ref{recon})] described in the previous section.

To compute the noise error $\Delta \bmath{C}_{\ell,{\rm noise}}$ we have to estimate the noise bias on the CMB power spectrum, say $\bmath{N}_{\rm CMB}$, i.e. the power spectrum of the noise in the reconstructed CMB map. This can be done with ``noise-only'' Monte Carlo simulations whereby several sets of realistic noise maps are generated and each of them is mixed with the matrix $\bmath{\sf W}$. Then the power spectrum is computed for each simulation and, finally, all the noise power spectra are averaged. 
Here we make the simplifying assumptions of Gaussian white noise and spatially invariant foreground spectral properties. This allows us to proceed analytically. In the framework of a realistic spatially-varying spectral model, we have verified that the analytic computation of the noise error with mean spectral dependencies over the sky is still enough accurate for the purpose of the forecast here considered. In fact, due to the intrinsic uncorrelation between noise and foregrounds, the mean level of the noise bias is predicted with good accuracy.

The theoretical power spectrum of Gaussian white noise at the frequency $\nu$ is:
%To compute the noise contribution we start from
%theoretical power spectrum of noise at the frequency $\nu$:
\begin{equation}\label{clnoise}
\bmath{N}_{\nu}=\frac{4 \pi {\rm fsky}}{\rm Npix} \sigma^2_{\nu}\bmath{B}_{\nu}^2
\end{equation}
where $\rm fsky$ is the sky fraction considered, $\rm Npix$ the number of pixels,  $\sigma^2_{\nu}$ the noise variance at frequency $\nu$ and $\bmath{B}_{\nu}^2(\ell)$  is the beam function applied to the channel maps to obtain a common resolution.  As in our case we deal with polarization,  $\sigma_{\nu}$ is the noise of the detector at frequency $\nu$ multiplied by $\sqrt{2}$ which, in the Gaussian white noise hipothesis here considered, gives the RMS of the noise in Q and U maps.

Given the linearity of the CMB recovery process [eq.~(\ref{recon})] and under the assumption of a spatially invariant reconstruction matrix, the noise bias is obtained by combining the channel noise spectra $\bmath{N}_{\nu}$ with the matrix $\bmath{\sf{W}}^2$:
\begin{equation}\label{clnoise}
\bmath{N}_{\rm CMB}=\sum_\nu w^2_{\nu,{\rm CMB}}\, \bmath{N}_{\nu},
\end{equation}
where $w^2_{\nu,{\rm CMB}}$ are the elements of the matrix $\bmath{\sf{W}}^2$ pertaining to the CMB component.

The error on the CMB power spectrum is due to the fact that we do not know the actual noise realization, but only a mean noise bias. In other words, $\Delta  \bmath{C}_{\ell,{\rm noise}}$  is the error due to the sampling variance of the noise bias $\bmath{N}_{\rm CMB}$:
% noise variance respect to this noise bias:
\begin{equation}
\label{deltanoise}
%\Delta  C_{\ell,{\rm noise}} =\sqrt{\frac{2/(2\ell+1)}{\rm fsky \, nbin}}\sum_{N_{ch}} (w_\nu^{CMB})^2 N_{\nu}(\ell),
\Delta  \bmath{C}_{\ell,{\rm noise}} =\sqrt{\frac{2/(2\ell+1)}{\rm fsky \, \bmath{nbin}(\ell)}}\,\bmath{N}_{\rm CMB},
\end{equation}
where $\bmath{nbin}$ is a function containing the number of multipoles around any chosen $\ell$ to be averaged according to a certain binning scheme.

The error $\Delta \bmath{C}_{\ell,{\rm foreg}}$ [eq.~(\ref{total})] due to the imperfect foreground subtraction can be computed following Stivoli et al. (2010). The map of residuals, $\bmath{s}-\bmath{\hat s}$, for a linear mixture source reconstruction can be estimated as:
\begin{equation}
\label{deltacomp}
\bmath{s}-\bmath{\hat s}=(\bmath{\sf{W}}\bmath{\sf{H}}-\bmath{\sf{I}})\,\bmath{\tilde s},
\end{equation}
where $\bmath{\sf{I}}$ is the identity matrix and  $\bmath{\tilde s}$ is a set of simulated components. $\Delta C_{\ell,{\rm foreg}}$ is the power spectrum of the residuals computed over a certain area of the sky. We note that in the computation of the foregorund residuals we do not proceed analytically up to the power spectrum level. This means that, in this case, we could easily exploit a realistic spatially-varying foreground spectral model. Given the lack of observational constraint at present, however, for the purpose of the paper we are relying again on the spatially-invariant model.
 
According to eq.~(\ref{deltacomp}) the foreground residuals vanish for $\bmath{\sf{W}}=\bmath{\sf{H}}^{-1}$, which is the exact solution of the component separation problem in the absence of noise. For the reconstruction matrix of eq.~(\ref{gls}), the foreground contamination is non null (although very low) even for $\bmath{\sf{\hat H}}=\bmath{\sf{H}}$. It obviously increases with increasing error in the mixing matrix estimation. In the MV case the mismatch between $\bmath{\sf{W}}$ and $\bmath{\sf{H^{-1}}}$ is higher, and the foreground contamination increases.

It is clear that the  estimate of $\Delta \bmath{C}_{\ell,{\rm foreg}}$ through the previous relation is model-dependent since it relies on simulations of the data $\bmath{\tilde s}$ which are hampered by our poor knowledge of polarized foregrounds. The situation will substantially improve in the near future as new polarization data, above all those from the {\sc Planck} mission, will become available. 

\section{Application to the {\sc Planck} and COrE experiments}\label{sec:quattro}
%In this section we apply our forecast to the specifications of the on-fight {\sc Planck} satellite and to a candidate configuration for the ESA B-Pol project, a next-generation satellite dedicated to B-modes detection.
The adopted experimental specifications for the {\sc Planck} and COrE experiments are reported in Table 1. {\sc Planck} specifications are derived from Mandolesi et al. (2010) for the Low Frequency Instrument (LFI) and from Lamarre et al. (2008) for the High Frequency Instrument (HFI) assuming a mission duration of 29 months. The specification for COrE are from the ESA M3 call (Dec 2010) available at \emph{http://www.core-mission.org}. %The specifications of both experiments are respectively in Table 1a and 1b.

%The specifications for COrE are from the baseline configuration assumed for the proposal submitted to ESA on 3rd December 2010 (\emph{http://www.b-pol.org/latest.html}).

For each experiment we computed the error on the power spectrum using eqs.~(\ref{total})--(\ref{deltacomp}) for both the minimum variance (MV) and component separation (CS) case.  %This will allow us in sec. \ref{sec:results} to prove the capabilities of our forecast and to present some results.

\subsection{Sky model}
The polarization Q and U synchrotron and dust templates of our model were obtained at 100 GHz by running the  Planck Sky Model (PSM). \footnote{http://www.apc.univ-paris7.fr/APC\_CS/Recherche/Adamis/PSM/psky-en.php}
%http://www.apc.univ-paris7.fr/APC_CS/Recherche/Adamis/PSM/psky-en.php}. 
Extrapolations to lower and higher frequencies were made using the spectra of eq.~(\ref{scaling_synchro}) with $\beta_s=3$ for synchrotron and of eq.~(\ref{scaling_dust}) with $\beta_d=1.7$ and $T_d=18\,$K for dust. Is has been necessary to slightly adjust the intensity of the synchrotron template to reproduce the WMAP 7-yr K-band polarization map.
%The simulated maps of the polarized synchrotron emission were obtained from the Haslam et al. (1995) map with geometrical suppression factor, polarization angles and polarization fractions based on the magnetic field model of Miville-Desch\^enes et al. (2008). The 23 GHz map obtained in this way was subsequently adjusted to reproduce the WMAP 7-yr K-band polarization map. Extrapolations to higher frequencies were made using the spectrum of eq.~(\ref{scaling_synchro}) with $\beta_s=3$.
%The polarized dust maps were simulated {\bf DA RIVEDERE starting from the one produced by the {\sc Planck} sky model, version ...\footnote{SITO PSM}, at 100 GHz (??). This map was extrapolated to the other frequencies using eq.~(\ref{scaling_dust}) with $\beta_d=1.7$ and $T_d=18\,$K.}
%
The polarized CMB simulation is based on a standard  $\Lambda CDM$ model with WMAP 7-yr  cosmological parameters (Larson et al. 2010), including gravitational lensing. We have added tensor modes with tensor to scalar ratio r=[0.1,0.03,0.01,0.001] and re-ionization optical depth  $\tau =0.1$.

\subsection{Minimum noise variance weighting}
As mentioned above, the MV approach assumes that, outside a suitable mask, foregrounds are negligible so that we need to deal only with CMB and noise. Therefore the reconstruction matrix, $\bmath{\sf{W}}$, contains only inverse noise variance weights for the CMB and null elements for the foreground components. For this assumption to be plausible we must restrict ourselves to a  ``CMB sensitive'' set of channels and adopt a wide Galactic mask. To compute $\bmath{\sf{H}}$ and $\bmath{\sf W}$ we have thus considered only the channels in the frequency range $70 \leq\nu<200$ GHz and computed  the power spectrum of the foreground residual map [eq.~ (\ref{deltacomp})] only over the high Galactic latitude sky [$|b|\ge 30^\circ$, so that $\rm{fsky}=0.5$, see  eq.~(\ref{deltanoise})].

Given the different sensitivities of {\sc Planck} and COrE we have used different bin sizes. Bin centers for $\ell \le 150$ are $\ell_{bin}=[4,9,14,20,28,39,53,80,140]$  for {\sc Planck} and $\ell_{bin}=[3,6,10,19,31,49,74,106,150]$ for COrE.
%For {\sc Planck} the bin size gradually increases from $\Delta \ell = 5$ for the lowest multipoles to $\Delta \ell = 40$ for $\ell\sim 100$. In the case of COrE $\Delta \ell$ increase from 3 to 25.

\subsection{Component separation weighting}
For the component separation approach we generated the estimated mixing matrix assuming the two standard polarized foreground components (synchrotron and thermal dust) with spatially invariant spectra, but allowing for errors in the estimate of synchrotron and dust spectral indices.
%Forecasting the latter is very difficult: it depends in general on the real sky (intensity, morphology and frequency scaling of the Galactic components), on the experiment (frequency, resolution and noise levels) and on the method exploited to estimate the mixing matrix.
Three error regimes, labeled as ``conservative'', ``intermediate'', and ``goal'' were considered (see Table~\ref{tab:rms}).

The ``conservative'' errors on spectral indices are those estimated by Ricciardi et al. (2010) applying the CCA component separation method (Bonaldi et al. 2006) to a realistic simulation of {\sc Planck} polarization data, including spatially-varying mixing matrix and realistic noise for 2 all-sky surveys. Thus they are slightly pessimistic for the real {\sc Planck} mission, now extended to 4 surveys. Moreover the next generation experiments targeted to B-modes detection will undoubtedly improve the accuracy of the determination of foreground spectral properties. That's why we also considered the ``intermediate'' and ``goal'' regimes.

\begin{table}
\centering
\caption{Adopted errors in the estimation of the synchrotron and dust spectral indices for three different regimes} \label{tab:rms}
\begin{tabular}{|l|l|l|}
%Detector&LFI&HFI\&\end{tabular}
\hline
&$\beta_s$& $\beta_d$\\
\hline
conservative&0.05&0.01\\
intermediate&0.01&0.005\\
goal&0.005&0.001\\
\hline
\end{tabular}
\end{table}

\begin{figure*}
\begin{center}
\includegraphics[width=8cm,keepaspectratio]{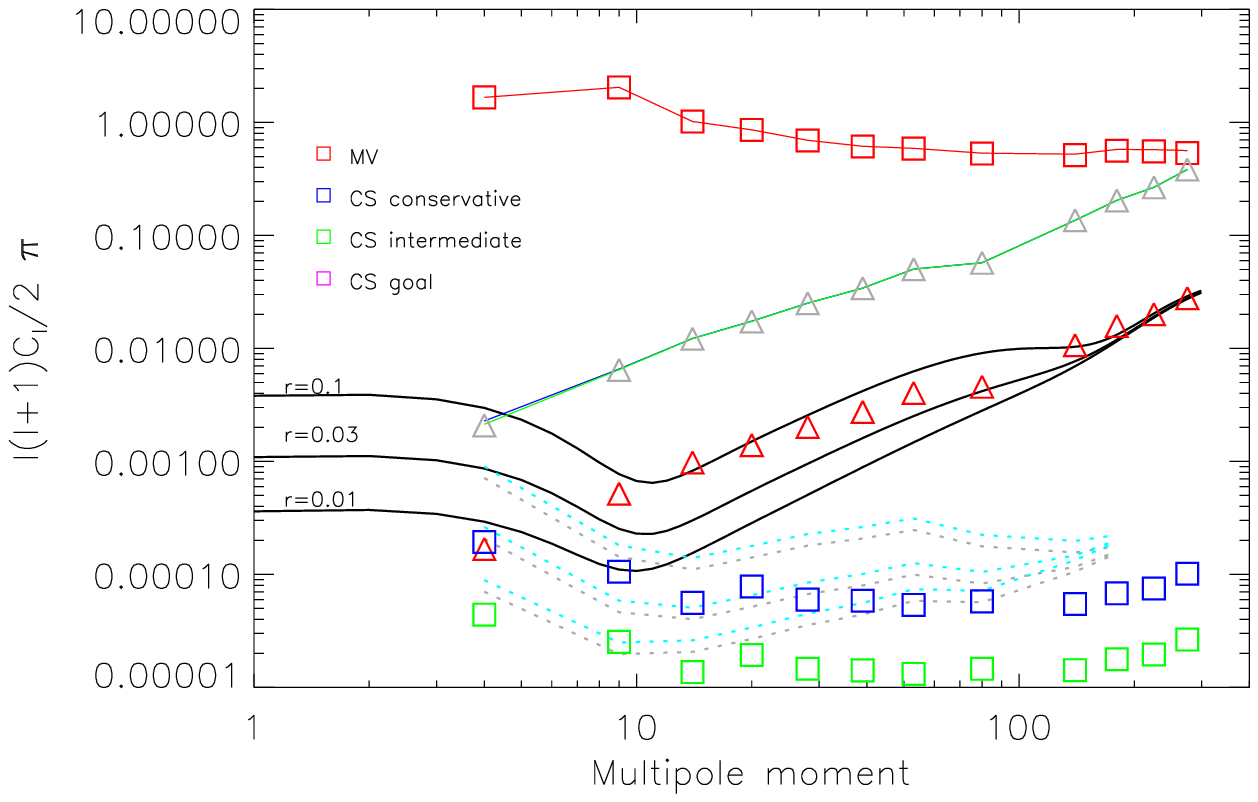}
\includegraphics[width=8cm,keepaspectratio]{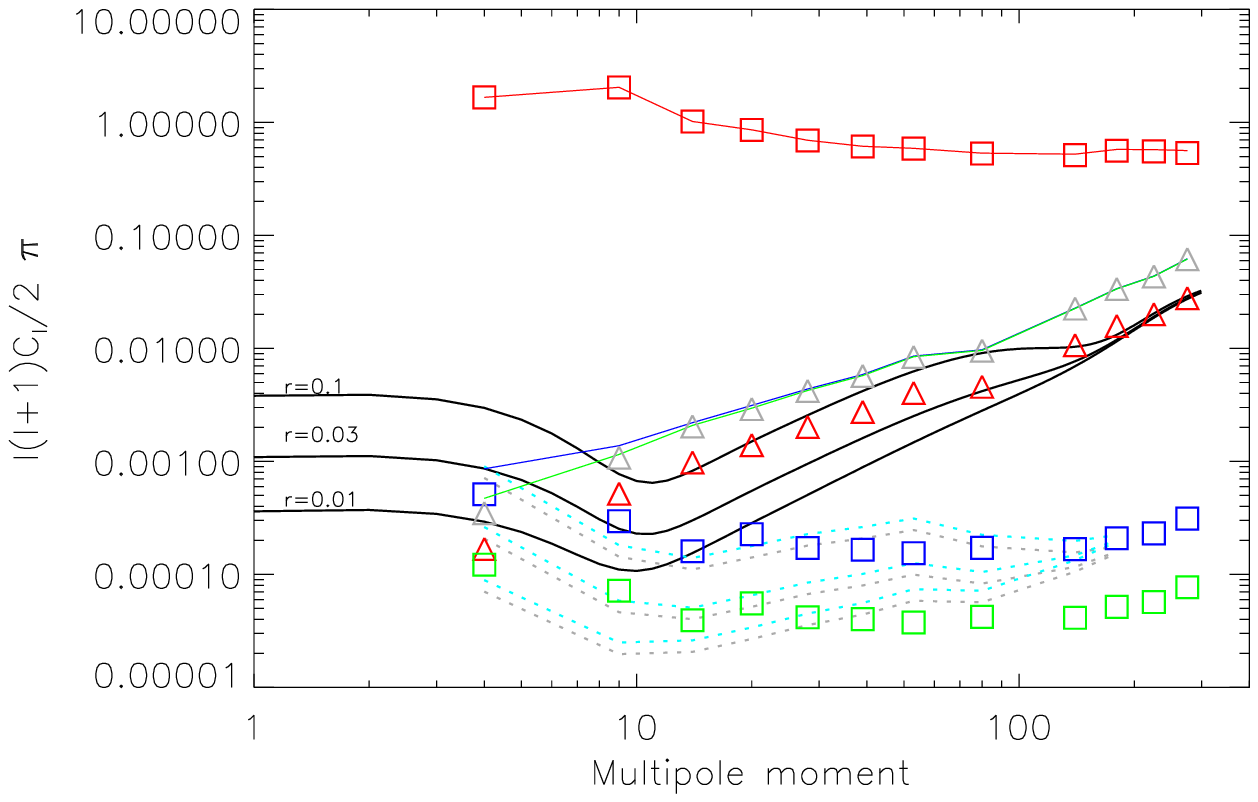}
\includegraphics[width=8cm,keepaspectratio]{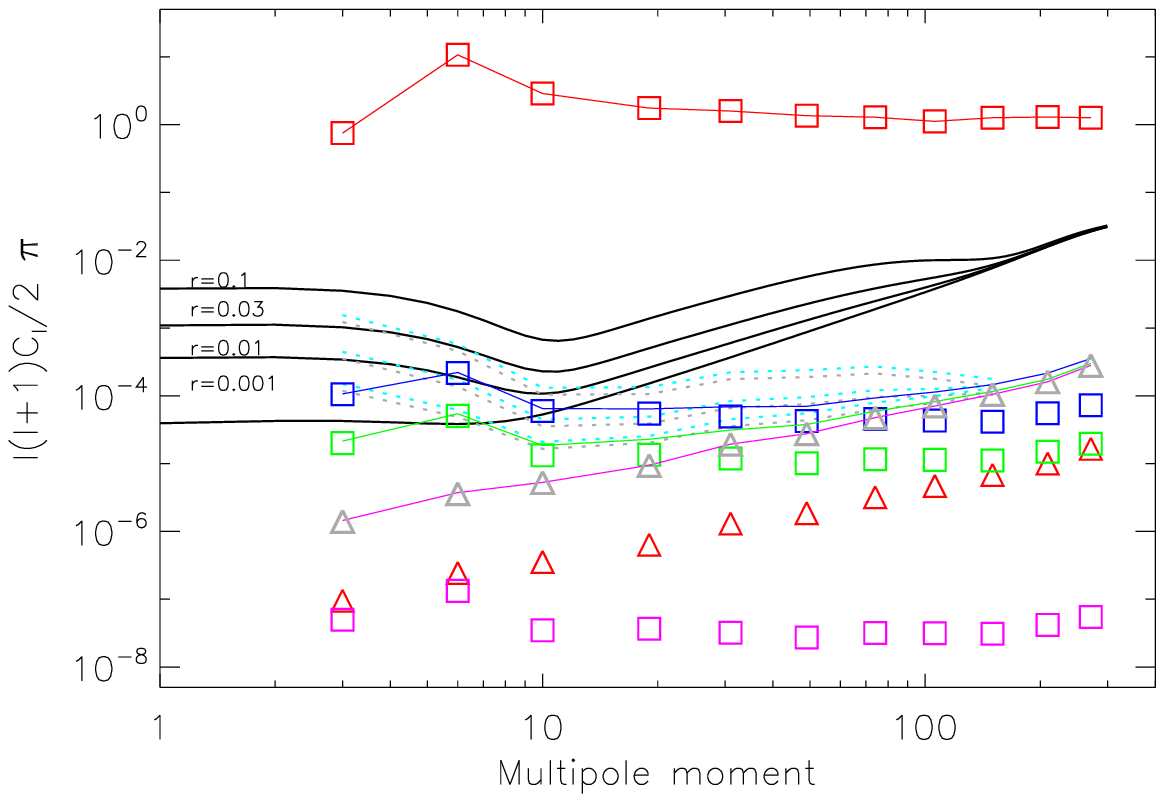}
\includegraphics[width=8cm,keepaspectratio]{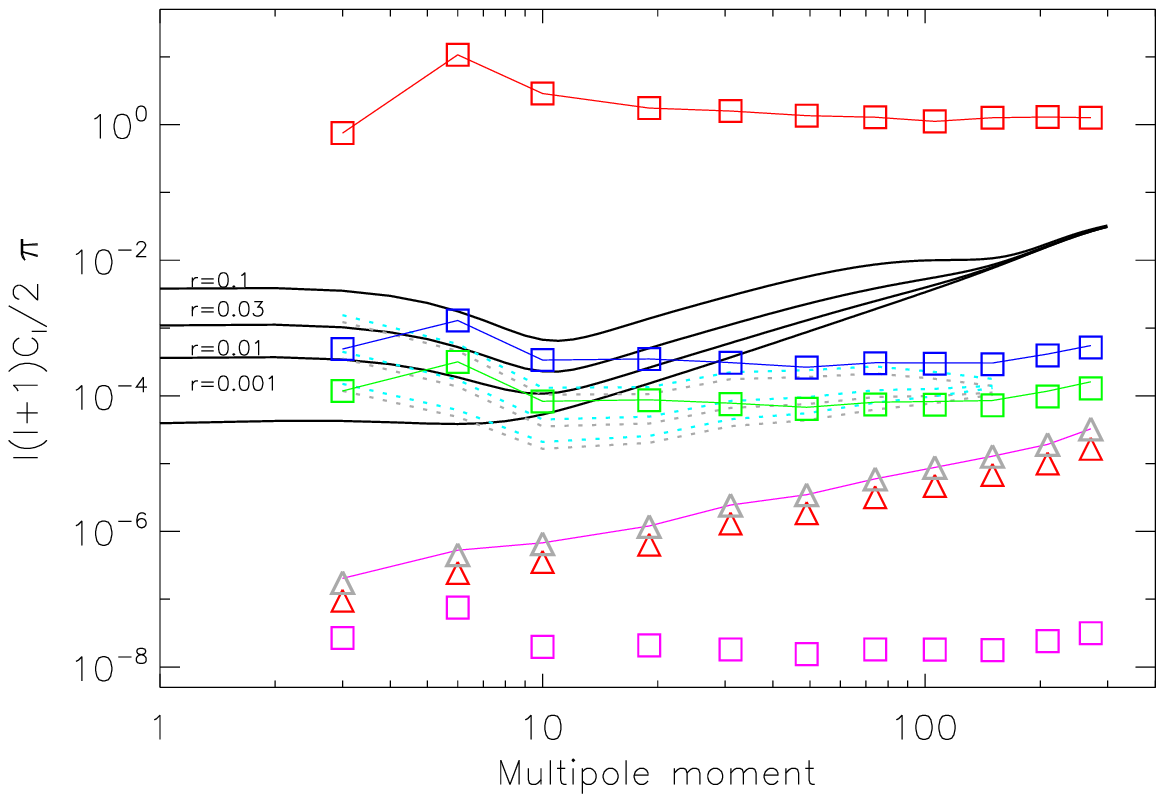}
\caption{Forecast for B-modes detection by {\sc Planck} (upper panels) and COrE (lower panel) for two choices of the frequency range used for the polarized CMB reconstruction: $50<\nu<200$ GHz (left-hand panels) and $40<\nu<300$ GHz  (right-hand panels). Black solid lines: fiducial models for tensor to scalar ratios, $r$, of 0.1, 0.03 0.01 and 0.001. Dotted lines: cosmic variance limits (cyan for 50\% sky coverage, grey for 85\%). Coloured solid lines: total error for MV (red) and CS with conservative (blue), intermediate (green) and goal (magenta) errors in the mixing matrix estimations (see Table~\protect\ref{tab:rms}). Those total errors are the sum of the noise (red triangles for MV, grey triangles for CS) and foreground (coloured squares) errors.}\label{planck}
\end{center}
\end{figure*}

To propagate the errors on spectral indices to the final result, we generated a set of 10 estimated mixing matrices $\bmath{\sf {\hat H}}$, drawing the actual spectral indices from  Gaussian distributions with $\sigma$ equal to the spectral index errors. For each  $\bmath{\sf {\hat H}}$ we computed the reconstruction matrix $\bmath{\sf W}$, the residual map through eq.~(\ref{deltacomp}) and its power spectrum. Since in this case we minimize foreground residuals the usable sky fraction is wider. We adopt   $\rm{fsky}=0.85$, corresponding to a Galactic cut of $\pm 10$ deg around the equator.
The values of $\Delta C_{\ell,{\rm foreg}}$ shown in Fig.~\ref{planck} are the mean for the 10 simulations.
%For the CS case we assumed the same bin function $\bmath{nbin}$ of the MV case.

Another issue is the choice of the set of channels to be used for the source reconstruction. On one hand, using a wide frequency range brings in channels heavily foreground contaminated and, since the foreground spectra are not perfectly known, this increases the foreground residuals in the final map. On the other hand, using a too limited set of channels in the linear combination leads to a higher noise level in the reconstructed CMB map, because the mixing matrix is less well-conditioned.

The best trade-off clearly depends on the instrument, and in particular on its sensitivity. To exemplify this situation we considered two sets of channels: a restricted one, limited to the frequency range  $50<\nu<200$ GHz, and a wider one  ($40<\nu<300$ GHz) for both the considered experiments.

We stress that this analysis on the choice of channels refers to the {\it source reconstruction}. For the estimation of the {\it mixing matrix}, i.e.  of the foreground spectral properties, it is useful to exploit a frequency range as wide as possible. To this end the lowest and highest frequencies are particularly useful, as they map respectively polarized synchrotron and dust emissions with high signal-to-noise and low contamination from the other components. Even if not explicitly addressed in this paper, the mixing matrix estimation has a key role in the detection of B-modes, as will be shown by the comparison of the results on the spectral index errors for our different regimes.

\section{Results} \label{sec:results}

In Fig.~\ref{planck} we show the result of the forecast for {\sc Planck} (upper panels) and COrE (lower panel) for the  two choices of the frequency range (left: $50<\nu<200$ GHz, right: $40<\nu<300$ GHz) used for the CMB map reconstruction. We can immediately note that the MV approach performs worse: the noise level is low, but the total error is strongly dominated by foregrounds even with a Galactic mask covering 50\% of the sky ($|b|\le 30^\circ$). 
To verify wether this result depends critically on the mask, we also combined the $|b|\le 30^\circ$ mask with the WMAP 7yrs polarization mask, finally exploiting 46\% of the sky. In this case, the foreground contamination lowers approximately by 30\% at large scales. With a mask optimized for the frequency coverage here considered we could do even better; nonetheless, the residual contamination would be still very high respect to the CMB. We conclude that on large and intermediate scales a proper component separation is necessary.

In the {\sc Planck} case, the total error for the CS approach is noise dominated even for a 85\% sky coverage ($|b| \ge 10^\circ$). Extending the frequency range from $50<\nu<200$ GHz to the $40<\nu<300$ GHz improves substantially the noise level. To be more quantitative, let's call $A^2$ the sum of the elements of the matrix $\bmath{\sf W}^2$ pertaining to the CMB component. These elements weight the noise power spectra in the computation of the noise bias $\bmath{N}_{\rm CMB}$. Thus $A$ can be interpreted as the amplification of the noise rms. In our simulation, the smaller set of channels yields $A \sim 3$, while the larger one yields $A \sim 1.3$. Our simulations indicate that, using the wider frequency range, {\sc Planck} can detect B-modes for $r=0.1$ and, with lower significance, for $r=0.03$. The fact that the error is dominated by the noise term implies that the improvement in the mixing matrix estimation from the ``conservative'' to the ``intermediate'' case has a minor effect.

Although our conclusions on {\sc Planck}''s capability for detecting B-modes are similar to those of  Efstathiou \& Gratton (2009), our approach differs substantially from theirs. These authors based their analysis on a minimum noise variance combination of the 70, 100 and 143 GHz channels and assumed that foregrounds can be removed to high accuracy if the highest and lowest polarized {\sc Planck} channels are used as foreground templates and the most contaminated areas (37\% of the sky) are masked. However, as shown by our analysis, the foreground contamination corresponding to the minimum noise variance combination is very high even at high Galactic latitude. Therefore a sufficiently accurate foreground removal by template fitting is very challenging. Conversely, a more standard component separation approach, as the one exploited for our forecast, keeps both foreground contamination and noise under control.

The bottom panels of Fig.~\ref{planck} show our expectations for the COrE mission. Due to the higher sensitivity of this experiment, the total error for the CS case is dominated by noise only for very low errors in the mixing matrix estimation (``goal''). For ``conservative'' errors in the spectral indices, the best results are obtained with the restricted set of channels. B-mode polarization can be detected down to $r=0.01$ even with the "conservative" errors in the spectral indices. The minimum value of $r$ that can be reached depends on our ability to estimate the mixing matrix. 
\section{Conclusions}
\label{sec:conclusions}
We presented a method to forecast the detectability of CMB B-mode polarization on large and intermediate angular scales given the experimental specifications of a full-sky multi-frequency experiment. Our forecast accounts for both noise and residual foreground contamination after the CMB has been extracted from the data by means of a suitable linear combination of the frequency maps. The computation of foreground residuals explicitly includes errors in the modeling of the data and in the estimation of the frequency scalings of the components. The forecast is quick and flexible, and allows the selection of different sets of frequencies and of different methods for the CMB reconstruction. In this paper we considered the minimum noise variance (MV) and GLS component separation (CS) methods, but other methods can be easily implemented.

We have investigated several issues: 

\begin{itemize}

\item How do the performances of minimum noise variance methods compare with methods exploiting component separation? We have shown that, although component separation leads to a noise amplification by an amount ultimately depending on the conditioning of the mixing matrix,  in the low multipole regime considered in this paper we cannot dispense from the component separation step, e.g. by resorting to an aggressive masking of foreground contaminated regions, because the foreground contamination is more severe than that of noise and the mask cannot be too extended because large areas are essential to restrain cosmic variance.

\item Which is the optimal frequency range for the reconstruction of CMB polarization maps? Using a wide frequency range brings in channels heavily foreground contaminated and, since the foreground spectra are not perfectly known, increases the foreground residuals in the final map. On the other hand, using a too limited set of channels in the linear combination leads to a higher noise level in the reconstructed CMB map, because the mixing matrix is less well-conditioned. We find that if the total error is dominated by noise, like in the case of {\sc Planck}, it is convenient to use a rather broad frequency range ($40<\nu<300$ GHz) while in the case of more sensitive experiments, like the planned COrE mission, it is preferable to restrict ourselves to the ``cleaner'' frequency range  $50<\nu<200$ GHz. We stress, however, that this applies to the CMB reconstruction step. For the preliminary ``mixing matrix estimation'' step a much broader frequency coverage is essential.
    
\item How critical is a better understanding of polarized diffuse foregrounds for measurements of the power spectrum of primordial B-mode polarization? We find that in the case of {\sc Planck} the main limitation comes from the detector noise. We estimate that with four all-sky surveys {\sc Planck} can detect the re-ionization bump of the B-mode power spectrum for values of the tensor to scalar ratio $r$ down to $<0.1$, in agreement with the earlier conclusion by Efstathiou \& Gratton (2009). On the other hand, to fully exploit the sensitivity of planned experiments specifically aimed at measuring the primordial B-mode anisotropies, it is essential to substantially improve the determination of spectral properties of polarized foregrounds. {\sc Planck} maps will allow an important step forward in this direction.

\end{itemize}

As soon as more accurate multi-frequency maps of polarized foreground emissions are available, our method will allow us to make more quantitative predictions that may help the design of next generation CMB polarization experiments and the optimization of the data analysis strategy.

\section{Acknowledgements}
Work supported by ASI through ASI/INAF Agreement I/072/09/0 for the Planck LFI Activity of Phase E2.

This research used resources of the National Energy Research Scientific Computing Center, which is supported by the Office of Science of the U.S. Department of Energy under Contract No. DE-AC02-05CH11231. 

We thank C. Baccigalupi, P. Natoli and G. Polenta for their contributions to our work and the {\sc Planck} and COrE teams for useful discussions. 
We acknowledge the anonymous referee for useful comments and suggestions leading to improvements in the paper.

\end{document}